# Influence of Interband Interaction on Isotope Effect Exponent of $MgB_2$ Superconductors

P.Udomsamuthirun[1], C.Kumvongsa[2], A.Burakorn[1], P.Changkanarth[1]

[1] Department of Physics, Faculty of Science, Srinakharinwirot University, Bangkok 10110, Thailand. E-mail: udomsamut55@yahoo.com

[2] Department of Basic Science, School of Science, The University of the Thai Chamber of Commerce, Dindaeng, Bangkok 10400, Thailand.

## Abstract

The exact formula of $T_c$'s equation and the isotope effect exponent of two-band s-wave superconductors in weak-coupling limit are derived by considering the influence of interband interaction. The paring interaction in each band consisted of 2 parts : the electron-phonon interaction and non-electron-phonon interaction are included in our model. The isotope effect exponent of $MgB_2$, $\alpha = 0.3$ with $T_c \approx 40$ K, can be found in the weak coupling regime and interband interaction of electron-phonon interaction show more effect on isotope effect exponent than of non-phonon interaction.

# 1. Introduction

The isotope effect exponent, $\alpha$, is one of the most interesting properties of superconductors. In the conventional BCS theory $\alpha = 0.5$ for all element.
In high-$T_c$ superconductors, experimenter found that $\alpha$ is smaller than 0.5[1-3]. This unusual small value leads to suggestion that the pairing interaction might be predominantly of electronic origin with a possible small phononnic contribution[4]. To explain the unusual isotope effect in high-$T_c$ superconductors, many models have been proposed such as the van Hove singularity[5-7], anharmonic phonon[8,9], pairing-breaking effect[10], and pseudogap[11,12].

The discovery of [13] of superconductivity in $MgB_2$ with a high critical temperature, $T_c \approx 39$ K, has attracted a lot of considerable attention. Various experiments [14-21] suggest the existence of mutiband in $MgB_2$ superconductors. The gap values $\Delta(k)$ cluster into two groups at low temperature, a small value of $\approx 2.5$ meV and a large value of of $\approx 7$ meV. The calculation of the electron structure [22-26] support this conclusion. The Fermi surface consists of four sheets : two three-dimensional sheets from the $\pi$ bonding and antibonding bands $(2p_z)$, and two nearly cylindrical sheets from the two-dimensional $\sigma$ band $(2p_{x,y})$ [24,27]. There is a large difference in the electron-phonon coupling on different Fermi surface sheet and this fact leads to multiband description of superconductivity. The average electron-phonon coupling strength is found to be small values[14-16]. Ummarino et al.[28] proposed that $MgB_2$ is the weak coupling two band phononic system where the Coulomb pseudopotential and the interchannel paring mechanism are key terms to interpret the superconductivity state. Garland[29] has shown that Coulomb potential in the d-orbitals of transition metal reduce the isotope exponent whereas sp-metals generally shown a nearly full isotope effect. So for sp-metal as $MgB_2$, the Coulomb effect could not be account to explain the reduced of isotope exponent.

Budko et al.[30] and Hinks et al.[31] measured the boron isotope exponent and estimated as $\alpha_B = 0.26 \pm 0.03$ and nearly zero magnesium isotope effect. The boron isotope exponent is closed to that obtained for the $YNi_2B_2C$ and $LuNi_2B_2C$ borocarbideds [32,33] where theoretical work[34] suggested that the phonons responsible for the superconductivity are high-frequency boron optical modes. This observation is consistent with a phonon-mediated BCS superconducting mechanism that boron phonon modes are playing an important role.

The theory of thermodynamic and transport properties of $MgB_2$ was made in the framework of the two band BCS model [35-43]. Zhitomirsky and Dao[44] derive the Ginzburg-Landau functional for two gap superconductors from the microscopic BCS model and then investigate the magnetic properties. The concept of multiband superconductors was first introduced by Suhl[45] and Moskalenke[46] in case of large disparity of the electron-phonon interaction for different Fermi-surface sheets.

The purpose of this paper is to derive the exact formula of $T_c$ 's equation and the isotope effect exponent of two-band superconductors in weak-coupling limit by considering the influence of interband interaction. The paring interaction in each band consisted of 2 parts : a attractive electron-phonon interaction and a attractive non-electron-phonon interaction are included in our model.

# 2. Model and calculation

The properties of $MgB_2$ suggest the two-band s-wave superconductors($\sigma$-band and $\pi$-band). And in each band, it may have two energy gaps. To recover this fact, we make the assumption that the paring interaction consists of 2 parts : a attractive electron-phonon interaction and a attractive non-electron-phonon interaction in σ-band and π-band , and the σ− π scattering of interband pairs. The Hamiltonian of the corresponding system is taken in the form

$$H = H_\pi + H_p + H_{p\pi} \quad (1)$$

where $H_\pi$, $H_p$ and $H_{p\pi}$ are the Hamiltonian of π band , σ band and interband respectively that

$$H_\pi = \sum_{k\sigma} \varepsilon_{k\sigma} \pi^+_{k\sigma} \pi_{k\sigma} - \sum_{kk'} V_{\pi\pi kk'} \pi^+_{k\uparrow} \pi^+_{-k\downarrow} \pi_{-k'\downarrow} \pi_{k'\uparrow} \quad (2.1)$$

$$H_p = \sum_{k\sigma} \varepsilon_{k\sigma} p^+_{k\sigma} p_{k\sigma} - \sum_{kk'} V_{ppkk'} p^+_{k\uparrow} p^+_{-k\downarrow} p_{-k'\downarrow} p_{k'\uparrow} \quad (2.2)$$

$$H_{p\pi} = -\sum_{kk'} V_{p\pi kk'} (p^+_{k\uparrow} p^+_{-k\downarrow} \pi_{-k'\downarrow} \pi_{k'\uparrow} + \pi^+_{k\uparrow} \pi^+_{-k\downarrow} p_{-k'\downarrow} p_{k'\uparrow}) \quad (2.3)$$

Here we use the standard meaning of parameters and $V_{\pi\pi kk'}$, $V_{ppkk'}$, $V_{p\pi kk'}$ are the attractive interaction potential in σ band and π band ,and interband respectively .

By performing a BCS mean field analysis of Eq.(1) and applying standard techniques, we obtain the gap equation as

$$\Delta_{pk} = -\sum_{k'} V_{ppkk'} \frac{\Delta_{pk'}}{2\sqrt{\varepsilon_{pk'}^2 + \Delta_{pk'}^2}} \tanh(\frac{\sqrt{\varepsilon_{pk'}^2 + \Delta_{pk'}^2}}{2T}) - \sum_{k'} V_{p\pi kk'} \frac{\Delta_{\pi k'}}{2\sqrt{\varepsilon_{\pi k'}^2 + \Delta_{\pi k'}^2}} \tanh(\frac{\sqrt{\varepsilon_{\pi k'}^2 + \Delta_{\pi k'}^2}}{2T})$$

(3.1)

$$\Delta_{\pi k} = -\sum_{k'} V_{\pi\pi kk'} \frac{\Delta_{\pi k'}}{2\sqrt{\varepsilon_{\pi k'}^2 + \Delta_{\pi k'}^2}} \tanh(\frac{\sqrt{\varepsilon_{\pi k'}^2 + \Delta_{\pi k'}^2}}{2T}) - \sum_{k'} V_{\pi pkk'} \frac{\Delta_{pk'}}{2\sqrt{\varepsilon_{pk'}^2 + \Delta_{pk'}^2}} \tanh(\frac{\sqrt{\varepsilon_{pk'}^2 + \Delta_{pk'}^2}}{2T})$$

(3.2)

In each band, the paring interaction consists of 2 parts[47,48] : an attractive electron-phonon interaction $V_{ph}$ and an attraction non-electron-phonon interaction $U_c$. $\omega_D$ and $\omega_c$ is the characteristic energy cutoff of the Debye phonon and non-phonon respectively. The interaction potential $V_{kk'}$ may be written as

$$V_{ikk'} = -V_{ph}^i - U_C^i \quad \text{for} \quad 0 < |\varepsilon| < \omega_D$$
$$= -U_c^i \quad \text{for} \quad \omega_D < |\varepsilon| < \omega_c \quad \text{and} \quad i = p, \pi, p\pi$$

For such as the interaction the superconducting order parameter can be written as

$$\Delta_{jk} = \Delta_{j1} \quad \text{for} \quad 0 < |\varepsilon| < \omega_D$$
$$= \Delta_{j2} \quad \text{for} \quad \omega_D < |\varepsilon| < \omega_c \text{ and } j = p, \pi$$

### 3. $T_C$ 's Equation

In this section, the exact formula of $T_C$ 's equation of two- band s-wave superconductors is derived . At $T = T_C$ and constant density of state $N(\varepsilon_k) = N(0)$, Eq.(3) become

$$\begin{pmatrix} \Delta_{\pi 1} \\ \Delta_{p1} \\ \Delta_{\pi 2} \\ \Delta_{p2} \end{pmatrix} = \begin{pmatrix} (\lambda_\pi + \mu_\pi)I_1 & (\lambda_{\pi p} + \mu_{\pi p})I_1 & \mu_\pi I_2 & \mu_{\pi p}I_2 \\ (\lambda_{\pi p} + \mu_{\pi p})I_1 & (\lambda_p + \mu_p)I_1 & \mu_{\pi p}I_2 & \mu_p I_2 \\ \mu_\pi I_1 & \mu_{\pi p}I_1 & \mu_\pi I_2 & \mu_{\pi p}I_2 \\ \mu_{\pi p}I_1 & \mu_p I_1 & \mu_{\pi p}I_2 & \mu_p I_2 \end{pmatrix} \begin{pmatrix} \Delta_{\pi 1} \\ \Delta_{p1} \\ \Delta_{\pi 2} \\ \Delta_{p2} \end{pmatrix} \quad (6)$$

Here
$$I_1 = \int_0^{\omega_D} d\varepsilon \frac{\tanh(\varepsilon / 2T_C)}{\varepsilon} \quad (7.1)$$

$$I_2 = \int_{\omega_D}^{\omega_C} d\varepsilon \frac{\tanh(\varepsilon / 2T_C)}{\varepsilon} \quad (7.2)$$

and

$$\lambda_\pi = N_\pi(0) V_{ph}^\pi, \lambda_p = N_p(0) V_{ph}^p, \lambda_{\pi p} = N_\pi(0) V_{ph}^{\pi p} = N_p(0) V_{ph}^{p\pi}$$

and
$$\mu_\pi = N_\pi(0) U_C^\pi, \mu_p = N_p(0) U_C^p, \mu_{\pi p} = N_\pi(0) U_C^{\pi p} = N_p(0) U_C^{p\pi}$$

are the coupling constants.

Solving the secular equation, the appropriate solution is,

$$I_1 = \frac{A}{B + \sqrt{C^2 - D}} \quad (8)$$

that

$$A = 2(-1 + I_2\mu_p)(-1 + I_2\mu_\pi) - 2I_2^2\mu_{\pi p}^2$$

$$B = \mu_p + \mu_\pi + 2I_2(-\mu_p\mu_\pi + \mu_{\pi p}^2) + (\lambda_p + \lambda_\pi)((-1 + I_2\mu_p)(-1 + I_2\mu_\pi) - I_2^2\mu_{\pi p}^2))$$

$$C = \lambda_p + \lambda_\pi + \mu_p - I_2\mu_p(\lambda_p + \lambda_\pi) + \mu_\pi(1 + I_2^2\mu_p(\lambda_p + \lambda_\pi) - I_2(\lambda_p + \lambda_\pi + 2\mu_p))$$
$$\quad - I_2\mu_{\pi p}^2(-2 + I_2(\lambda_p + \lambda_\pi))$$

$$D = 4((-1 + I_2\mu_p)(-1 + I_2\mu_\pi) - I_2^2\mu_{\pi p}^2)[\lambda_\pi\mu_p - 2\lambda_{\pi p}\mu_{\pi p} - (-1 + I_2\lambda_\pi)(\mu_p\mu_\pi - \mu_{\pi p}^2)$$
$$\quad + \lambda_p[(-1 + I_2\mu_p)(-\mu_\pi + \lambda_\pi(-1 + I_2\mu_\pi)) - I_2\mu_{\pi p}^2(-1 + I_2\lambda_\pi)]$$
$$\quad + \lambda_{\pi p}^2(-1 + I_2(\mu_p + \mu_\pi) + I_2^2(-\mu_p\mu_\pi + \mu_{\pi p}^2))]$$

Eq.(8) is the $T_c$'s equation of two-band s-wave superconductors.

## 4. The isotope effect exponent

In harmonic approximation, $\omega_D \alpha\, M^{-1/2}$, and $\omega_c$ does not depend on mass. The isotope effect exponent can be derived from the equation

$$\alpha = -\frac{d \ln T_c}{d \ln M}$$
$$= \frac{1}{2} \frac{\omega_D}{T_c} \frac{dT_c}{d\omega_D} \quad (9)$$

where M is the mass of the atom constituting the specimen under consideration.

Using Eq.(6) and Eq.(9), we can the isotope effect exponent as below

$$\alpha = \frac{(1/2)}{\frac{\tanh(\omega_c / 2T_c)}{\tanh((\omega_D / 2T_c))} \frac{(\mu_\pi D' + \mu_p E' + \mu_{\pi p}^2 F')}{(\lambda_\pi A' + I_1\lambda_{\pi p}B' + \lambda_p C')} - 1} \quad (10)$$

Here
$$A' = -[-1 + \mu_p(I_1 + I_2)][-1 + \mu_\pi(I_1 - I_2)] + (I_1^2 - I_2^2)\mu_{\pi p}^2$$
$$B' = \lambda_{\pi p}[2(-1 + I_2\mu_p)(-1 + I_2\mu_p) + I_1(\mu_p + \mu_\pi - 2I_2\mu_p\mu_\pi)] + 4\mu_{\pi p} + 2(I_1 - I_2)I_2\lambda_{\pi p}\mu_{\pi p}^2$$
$$C' = (-1 + I_2\mu_p)(-1 + I_2\mu_\pi) + I_2^2\mu_{\pi p}^2 + I_1^2[-\mu_\pi + \mu_p(-1 + 2I_2\mu_\pi) - 2I_2\mu_{\pi p}^2]$$
$$\quad + I_1[\mu_p - \mu_\pi + 2\lambda_\pi(-(-1 + I_2\mu_p)(-1 + I_2\mu_\pi) + I_2^2\mu_{\pi p}^2)]$$
$$D' = I_1^2\lambda_{\pi p}^2 - (-1 + I_1\lambda_p)(-1 + I_1\lambda_\pi)$$
$$E' = -1 + I_1(\lambda_\pi + \lambda_p(1 - I_1\lambda_\pi) + I_1\lambda_{\pi p}^2) + \mu_\pi F$$
$$F' = 2I_2 + I_1(2 - 2I_2(\lambda_p + \lambda_\pi) + I_1(-\lambda_\pi + \lambda_p(-1 + 2I_2\lambda_\pi) - 2I_2\lambda_{\pi p}^2)$$

Eq.(8) and Eq.(10) can be easily reduced to the $T_c$'s equation and isotope exponent of BCS theory .

In Figure.(1), we plot a three dimensional graph of the isotope exponent Eq.(10) versus the interband coupling constant $\lambda_{\pi p}$ and $\mu_{\pi p}$. Depending on the measured Debye frequency $\omega_D$ =64.3 meV [30,49] and $T_c \approx 40$ K , the parameters are $T_c = 40$ K, $\omega_D = 745$ K , $\omega_c = 1.5\omega_D$, $\lambda_\pi = \lambda_p = 0.05$, $\mu_\pi = \mu_p = 0.05$. The isotope effect exponent is tend to 0.5 at large values of phonon and low value of non-phonon interband coupling constant . We calculate Eq.(8) and Eq.(10) numerically to find isotope effect exponent of MgB$_2$, $\alpha = 0.3$ with $T_c \approx 40$ K that many ranges of coupling constant agree with these conditions, example as $\mu_\pi = \mu_p = 0.05$ , $\lambda_{\pi p} = 0.05$, $\mu_{\pi p} = 0.142$, $0.034 < \lambda_p < 0.114$, and $0.01 < \lambda_\pi < 0.1$. In Figure.(2), we show the effect of interband coupling constant on isotope effect exponent . The interband interaction of electron-phonon interaction show more effect on isotope exponent than of non-phonon interaction and both of them increase the isotope effect exponent in the same way .

**5.Conclusions**

The exact formula of $T_c$'s equation and the isotope effect exponent of two-band s-wave superconductors in weak-coupling limit are derived by considering the influence of interband interaction . The paring interaction in each band consisted of 2 parts : a attractive electron-phonon interaction and a attractive non-electron-phonon interaction are included .We find isotope effect exponent of MgB$_2$, $\alpha = 0.3$ with $T_c \approx 40$ K in many ranges of coupling constant. These strength values of the coupling parameters indicate that the MgB$_2$ superconductor is in the weak coupling regime. The interband interaction of electron-phonon interaction show more effect on isotope exponent than of non-phonon interaction.

**Acknowledgement** The author would like to thank Thailand Research Fund for financial support and the university of the Thai Chamber of Commerce for partial financial support and the useful discussions from Prof.Dr.Suthat Yoksan.

**6.Refference**
1. B.Batlogg et al., Phys.Rev.Lett. **58**, 2333(1987).
2. D.E.Moris et al., Phys.Rev. B **37**, 5936(1988).
3. S.Hoen et al., Phys.Rev.B **39**,2269(1989).
4. F.Marsiglio, R.Akis, and J.P.Carbotte, Solid State Commun. **64**,905(1987).
5. J.Labbe and J.Bok, Europhys.Lett. **3**,1225(1987)
6. C.C.Tsuei,D.M.Newns,C.C.Chi, and P.C.Pattnaik, Phys.Rev.Lett. **65**,2724(1990).
7. R.J.Radtke and M.R.Norman, Phys.Rev. B **50**,9554(1994).
8. H.B.Schuttler and C.H.Pao, Phys.Rev.Lett. **75**,4504(1995).
9. L.Pietronero and S.Strassler, Europhys.Lett. **18**, 627(1992).
10. J.P.Carbotte, M.Greeson, and A.Perez-gonzalez, Phys.Rev.Lett. **66**,1789(1991).
11. T.Dahm, Phys.Rev.B **61**, 6381(2000).
12. P.Udomsamuthirun, Phys.Stat.Sol.(b) **226**,315(2001).
13 J.Nagamatsu,N.Nakagawa,T.Muranaka,Y.Zenitani and J.Akimutsu, Nature(London) **410**,63(2001).
14 F.Bouquet, Y.Wang,R.A.Fisher et al., Europhys,Lett. **56**,856(2001).
15.F.Bouquet,R.A.Fisher,N.E.Phillips et al., Phys.Rev.Lett. **87**,047001(2001).
16.Y.Wang,T.Plackowski, and A.Junod, Physica C **355**,179(2001).
17.F.Bouquet,Y.Wang,I.Sheikin et al., Physica C **385**,192(2003).
18.P.Szabo et al.,Phys.Rev.Lett. **87**,137005(2001).
19.X.K.Chen et al.,Phys.Rev.Lett. **87**,157002(2001).
20.S.Tsuda et al., Phys.Rev.Lett. **87**,177006(2001).
21.F.Giubileo et al.,Phys.Rev.Lett. **87**,177008(2001).
22.J.M.An and W.E.Pickett,Phys.Rev.Lett. **86**,4366(2001).
23.J.Kortus et al.,Phys.Rev.Lett. **86**,4656(2001).
24.A.Y.Liu,I.I.Mazin, and J.Kortus, Phys.Rev.Lett. **87**,087005(2001).
25.N.I.Medvedeva et al. Phys.Rev. B **64**,020502(2001).
26 I.I.Mazin and V.P.Antropo, Physica C **385**,49(2003).
27.Y.Kong et al.,Phys.Rev. B **64**, 020501(2001).
28 .G.A.Ummario,R.S.Gonnelli,S.Massidda, and A.Bianconi,cond-mat/0310284v1
29 J.W.Garland,Phys.Rev.Lett. **11**,114(1963).
30 S.L.Budko,G.Lapertot,C.Petrovic et al., Phys.Rev.Lett. **86**,1877(2001).
31 D.G.Hinks,H.Claus and J.D.Jorgensen, Nature **411**,457(2001).
32 D.D.Lawrie and J.P.Franck, Physica(Amsterdam) **245C**, 159(1995).
33 K.O.Cheon,I.R.Fisher, and P.C.Canfield, Physica(Amsterdam) **312C**, 35(1999).
34 L.F.Mattheiss,T.Siegrist, and R.J.Cava, Solid State Commun. **91**, 587(1994).
35 A.A.Golubov et al.,J.Phys.:Condens.Matter **14**,1353(2002).
36 A.Brinkman et al.,Phys.Rev.B **65**,180517(2002).
37 I.I.Mazin et al.,Phys.Rev.Lett. **89**,1072002(2002).
38 A.Nakai, M.Ichioka, and K.Machida, J.Phys.Soc.Jpn. **71**,23(2002).
39 P.Miranovic, K.Machida, and V.G.Kogan, J.Phys.Soc.Jpn. **72**,221(2003).
40 T.Dahm and N.Scopohl, Phys.Rev.Lett. **91**,017001(2003).
41 A.Gurevich, Phys.Rev. B **67**,184515(2003).
42 A.A.Golubov and A.E.Koshelev, cond-mat/0303237
43 A.E.Koshelev and A.A.Golubov, Phys.Rev.Lett. **90**,177002(2003).

44 M.E. Zhitomirsky and V.H.Dao, Phys.Rev.B **69**,054508(2004).
45 H.Suhl,B.T.Matthias, and L.R.Walker, Phys.Rev.Lett. **3**,552(1959).
46 V.A.Moskalenko, M.E.Palistrant, and V.M.Vakalyuk, Sov.Phys.Usp. **34**,717(1991).
47 S.Yoksan Solid State Commun. **78**, 233(1991).
48 L.L.Daemen and A.W.Overhauser, Phys.Rev. B**41**, 7182(1990)
49 Ch.Walti et al.,Phys.Rev. B **65**,172515(2001).

**Figure Caption**

**Figure(1).** Plot graph of the isotope exponent Eq.(10) versus the interband coupling constant $\lambda_{\pi p}$ and $\mu_{\pi p}$, the parameters are $T_c = 40$ K, $\omega_D = 745$ K, $\omega_c = 1.5\omega_D$, $\lambda_\pi = \lambda_p = 0.05$, $\mu_\pi = \mu_p = 0.05$.

**Figure.(2).** We show the effect of interband coupling constant on isotope effect exponent. The parameters are $T_c = 40$ K, $\omega_D = 745$ K, $\omega_c = 1.5\omega_D$, $\lambda_p = 0.1$, $\mu_\pi = \mu_p = 0.05$.

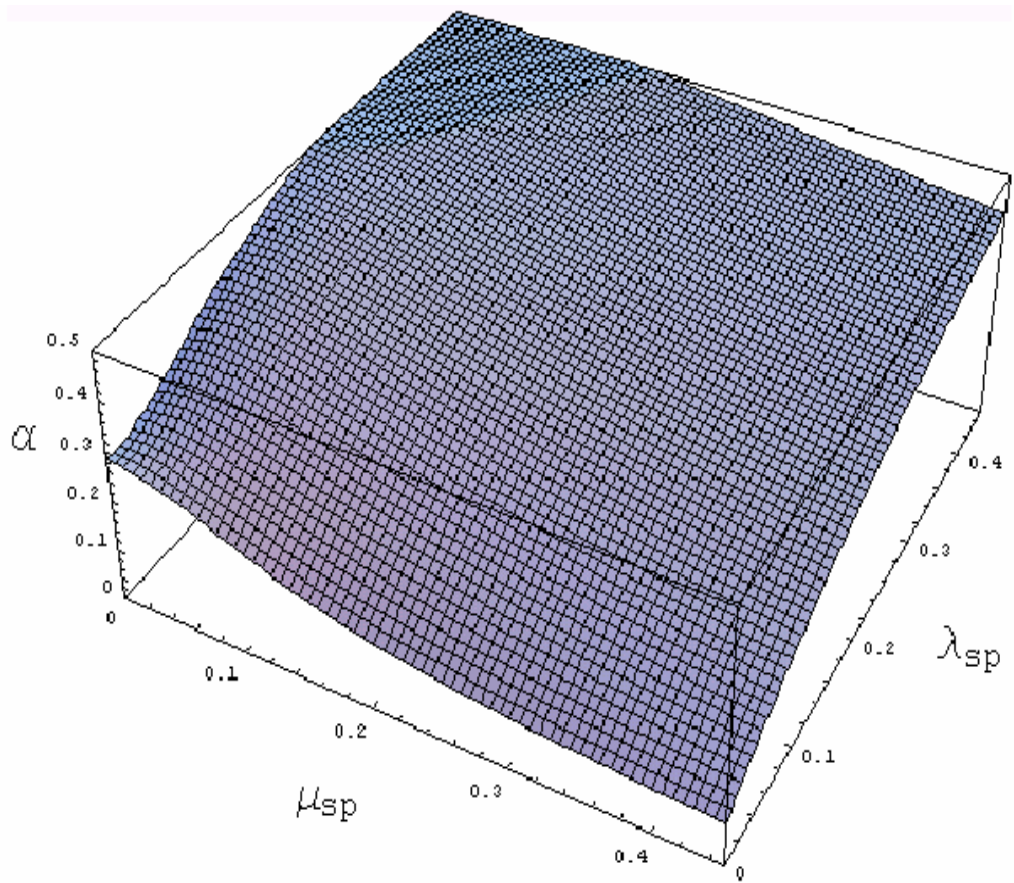

**Figure(1).**

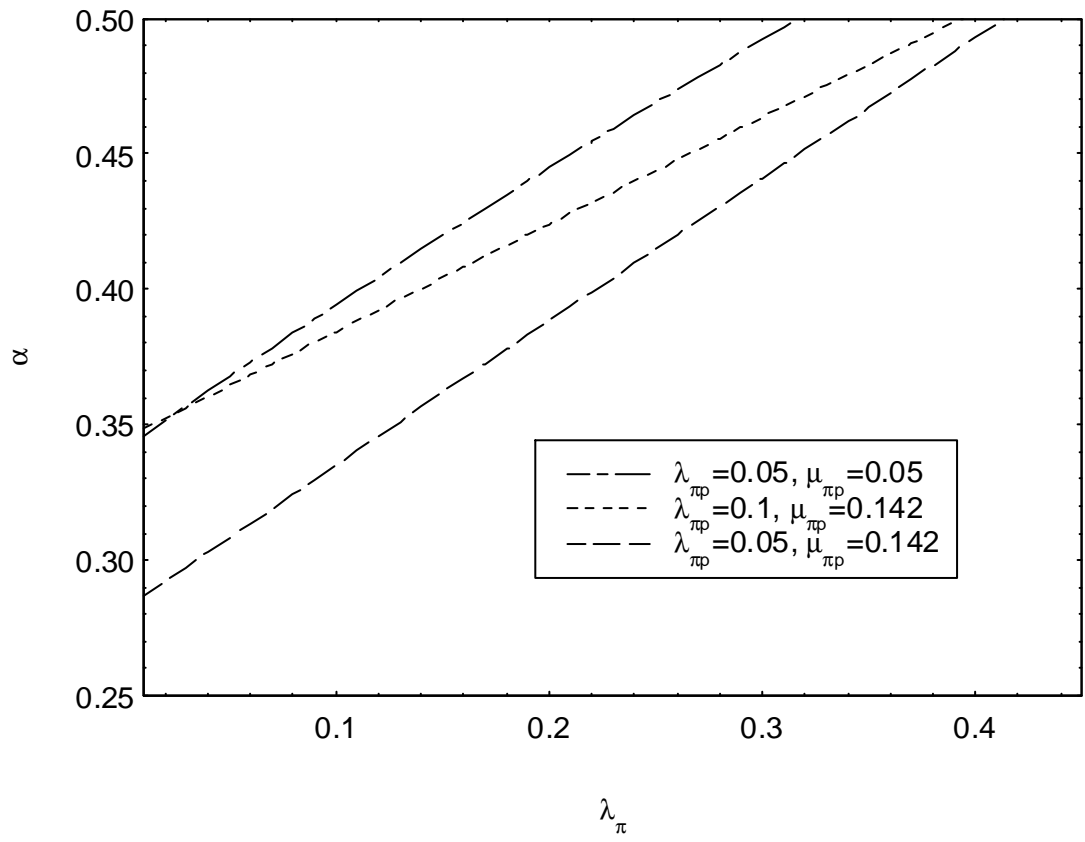

**Figure(2).**

Udomsamuthirun et al.